\documentclass[aps,prl,superscriptaddress,preprintnumbers,twocolumn,groupedaddress,showpacs]{revtex4}

\usepackage{graphicx}

\newcommand{\rT}{{\mathrm{T}}}

\def\mathswitchr#1{\relax\ifmmode{\mathrm{#1}}\else$\mathrm{#1}$\fi}
\newcommand{\PZ}{\mathswitchr Z}
\newcommand{\PH}{\mathswitchr H}
\newcommand{\Pb}{\mathswitchr b}
\newcommand{\Pep}{\mathswitchr {e^+}}
\newcommand{\Pem}{\mathswitchr {e^-}}
\newcommand{\Pp}{\mathswitchr p}
\newcommand{\Pg}{\mathswitchr g}
\newcommand{\Pt}{\mathswitchr t}
\newcommand{\Pu}{\mathswitchr u}
\newcommand{\Pd}{\mathswitchr d}
\newcommand{\PW}{\mathswitchr W}

\def\mathswitch#1{\relax\ifmmode#1\else$#1$\fi}

\newcommand{\Mt}{\mathswitch {m_\Pt}}

\newcommand{\TeV}{\unskip\,\mathrm{TeV}}
\newcommand{\GeV}{\unskip\,\mathrm{GeV}}
\newcommand{\MeV}{\unskip\,\mathrm{MeV}}

\newcommand{\fb}{\unskip\,\mathrm{fb}}

\def\reffi#1{\mbox{Figure~\ref{#1}}}

\def\citere#1{\mbox{Ref.~\cite{#1}}}
\def\citeres#1{\mbox{Refs.~\cite{#1}}}

\begin{document}

\preprint{MPP-2009-53, CERN-PH-TH/2009-055, PSI-PR-09-04}
\title{\boldmath{
NLO QCD corrections to $\Pp\Pp\to\Pt\bar\Pt\Pb\bar\Pb+X$ at the LHC}}
 
\author{A.~Bredenstein}
\affiliation{High Energy Accelerator Research Organization (KEK),
Tsukuba, Ibaraki 305-0801, Japan}

\author{A.~Denner}
\affiliation{Paul Scherrer Institut, W\"urenlingen und Villigen,
CH-5232 Villigen PSI, Switzerland}

\author{S.~Dittmaier}
\affiliation{Albert-Ludwigs-Universit\"at Freiburg, Physikalisches Institut, 
D-79104 Freiburg, Germany}
\affiliation{Max-Planck-Institut f\"ur Physik
(Werner-Heisenberg-Institut), D-80805 M\"unchen, Germany}

\author{S.~Pozzorini}
\affiliation{CERN, CH-1211 Geneva 23, Switzerland}

\date{\today}

\begin{abstract}
We report on the calculation of the full next-to-leading order QCD corrections
to the production of $\Pt\bar\Pt\Pb\bar\Pb$ final states at the LHC, which
deliver a serious background contribution to the production of a Higgs
boson (decaying into a $\Pb\bar\Pb$ pair) in association with a
$\Pt\bar\Pt$ pair. While the corrections significantly reduce the
unphysical scale dependence of the leading-order cross section, our
results predict an enhancement of the $\Pt\bar\Pt\Pb\bar\Pb$
production cross section by a $K$-factor of about $1.8$.
\end{abstract}


\maketitle


Extending earlier work~\cite{Bredenstein:2008zb}, where 
we discussed the next-to-leading order (NLO) QCD corrections
to $\Pt\bar\Pt\Pb\bar\Pb$ production via
quark--antiquark annihilation, in this letter we present first
results on the full NLO QCD corrections to $\Pp\Pp\to\Pt\bar\Pt\Pb\bar\Pb+X$
at the LHC, i.e.\ we complete the existing results by the contributions from
gluonic initial states.

The QCD-initiated production of $\Pt\bar\Pt\Pb\bar\Pb$ final states
represents a very important background to $\Pt\bar\Pt\PH$ production
where the Higgs boson decays into a $\Pb\bar\Pb$ pair.  While early
studies of $\Pt\bar\Pt\PH$ production at ATLAS~\cite{atlas-cms-tdrs}
and CMS~\cite{Drollinger:2001ym} suggested even discovery potential of
this process for a light Higgs boson, more recent
analyses~\cite{Cammin:2003,Cucciarelli:2006,Benedetti:2007sn,Aad:2009wy}
with more realistic background assessments show that the signal
significance is jeopardized if the background from
$\Pt\bar\Pt\Pb\bar\Pb$ and $\Pt\bar\Pt+\mathrm{jets}$ final states is
not controlled very well.  The calculation presented in this letter
renders improved signal and background studies possible that are based
on NLO predictions for $\Pt\bar\Pt\Pb\bar\Pb$ final states.  NLO QCD
corrections are already available for the $\Pt\bar\Pt\PH$ signal
\cite{Beenakker:2001rj,Beenakker:2002nc,Dawson:2002tg,Dawson:2003zu}
and the background from $\Pt\bar\Pt+\mathrm{jet}$
\cite{Dittmaier:2007wz,Dittmaier:2008uj} and $\Pt\bar\Pt\PZ$
\cite{Lazopoulos:2008de}.

On the theoretical side, the calculation of NLO corrections to $2\to4$
particle processes represents the current technical frontier. The
complexity of such calculations triggered the creation of prioritized
experimenters' wishlists~\cite{Buttar:2006zd,Bern:2008ef} for missing
NLO calculations for LHC physics, and the process of
$\Pt\bar\Pt\Pb\bar\Pb$ production ranges among the most wanted
candidates.  In recent years the field of NLO calculations to
multi-particle processes received an enormous boost, most notably by
advanced methods for evaluating one-loop tensor integrals for Feynman
diagrams~\cite{Ferroglia:2002mz,Denner:2002ii,Giele:2004iy,%
  Giele:2004ub,Binoth:2005ff,Denner:2005nn,Binoth:2008uq} and by new
methods employing unitarity cuts of one-loop amplitudes analytically
(see, e.g., \citere{Bern:2007dw} and references therein) or
numerically~\cite{Ossola:2006us,Ellis:2007br,
  Ossola:2007ax,Giele:2008ve,Berger:2008sj,Giele:2008bc}.  The
numerical unitarity-based approaches have successfully passed their
proof of principle in the calculation of specific one-loop QCD
amplitudes, including in particular multi-gluon
amplitudes~\cite{Giele:2008bc,Lazopoulos:2008ex,Winter:2009kd},
$\Pu\bar \Pd\to\PW^+q\bar q\Pg$~\cite{Ellis:2008qc}, $\Pu\bar
\Pd\to\PW^+\Pg\Pg\Pg$~\cite{Ellis:2008qc,vanHameren:2009dr}, $\Pu\bar
\Pu/\Pg\Pg\to\Pt\bar\Pt\Pb\bar\Pb$, $\Pu\bar
\Pu\to\PW^+\PW^-\Pb\bar\Pb$, $\Pu\bar \Pu\to\Pb\bar\Pb\Pb\bar\Pb$,
$\Pu\bar \Pu/\Pg\Pg\to\Pt\bar\Pt\Pg\Pg$~\cite{vanHameren:2009dr}, and
in the evaluation of the leading-colour contribution at NLO to W+3jet
production at the
Tevatron~\cite{Ellis:2008qc,Ellis:2009zw,Berger:2009zg}.  The
Feynman-diagrammatic approach has already been used in complete NLO
predictions for some $2\to4$ reactions at
$\Pep\Pem$~\cite{Denner:2005es,Denner:2005fg} and $\gamma\gamma$
colliders~\cite{Lei:2007rv} and in the evaluation of the amplitude of
$q\bar q\to\Pb\bar\Pb\Pb\bar\Pb$~\cite{Reiter:2009dk} at one loop.
The evaluation of $\Pt\bar\Pt\Pb\bar\Pb$ production---also based on
Feynman diagrams, as documented in \citere{Bredenstein:2008zb} and
this letter---represents the first full NLO calculation for a $2\to4$
process at a hadron collider.

\begin{figure}
{\includegraphics[bb=125 475 390 715, width=.38\textwidth]{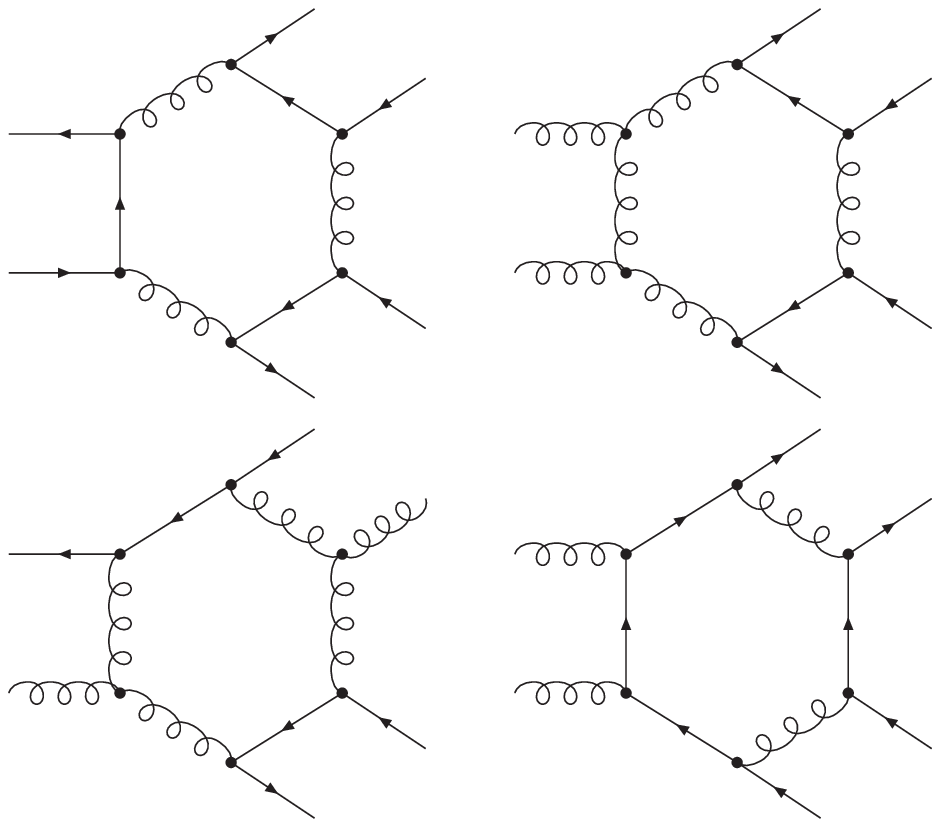}}
\vspace*{-1em}
\caption{Some generic hexagon diagrams for hadronic
$\Pt\bar\Pt\Pb\bar\Pb$ production at NLO QCD.}
\label{fig:hexagons}
\end{figure}

In LO QCD, 7 and 36 different Feynman diagrams contribute to the
production of $\Pt\bar\Pt\Pb\bar\Pb$ final states via
$q\bar q$ annihilation and $\Pg\Pg$ fusion, respectively.
The virtual QCD corrections comprise about 200 one-loop diagrams for the
$q\bar q$ and about 1000 diagrams for the $\Pg\Pg$ initial state, the
most complicated being the 8 and 40~hexagons for the respective channels.
Some hexagon graphs are depicted in \reffi{fig:hexagons}.
The real QCD corrections comprise gluon bremsstrahlung in the
$q\bar q$ and $\Pg\Pg$ channels, 
$q\bar q/\Pg\Pg\to\Pt\bar\Pt\Pb\bar\Pb\Pg$, 
and (anti)quark--gluon scattering processes,
$\raisebox{.6em}{\tiny $(-)$}\hspace{-.83em}q\,\Pg
\to\Pt\bar\Pt\Pb\bar\Pb \raisebox{.6em}{\tiny $(-)$}\hspace{-.83em}q$.
We consistently neglect contributions involving b~quarks in the
initial state because of their suppression in the parton distribution
functions (PDFs).
In the following we briefly describe the calculation of the virtual
and real corrections. Each of these contributions has been
worked out twice and independently, resulting in two completely
independent computer codes.

The evaluation of the virtual corrections starts with the generation
of the one-loop amplitudes via two independent versions of {\sc
  FeynArts}~\cite{Kublbeck:1990xc,Hahn:2000kx}.  Using either inhouse
{\sc Mathematica} routines or {\sc FormCalc}~\cite{Hahn:1998yk}, each
diagram is decomposed in terms of standard spin and colour structures,
as described in \citere{Bredenstein:2008zb} for the $q\bar q$ channel
in detail. The coefficients in the resulting linear combination of
these standard structures contain the one-loop tensor integrals.  The
obtained expressions are not reduced to standard scalar integrals
analytically, but the tensor integrals are evaluated by means of
algorithms that perform a recursive reduction to master integrals in
numerical form. This avoids a further increase of the huge analytic
expressions and permits to adapt the reduction strategy to the
specific numerical problems that appear in different phase-space
regions.  In detail, the 6-/5-point integrals are directly expressed
in terms of 5-/4-point integrals \cite{Denner:2002ii,Denner:2005nn}.
Tensor 4-point and 3-point integrals are reduced to standard scalar
integrals with the Passarino--Veltman
algorithm~\cite{Passarino:1978jh} as long as no small Gram determinant
appears in the reduction. If small Gram determinants occur, the
alternative schemes of \citere{Denner:2005nn} are applied.
Ultraviolet (UV) divergences are regularized dimensionally throughout,
but infrared (IR) divergences are treated in different variants, which
comprise pure dimensional regularization with strictly massless light
quarks (including b quarks) and a hybrid scheme with small quark masses.
The corresponding scalar master integrals are evaluated using the
methods and results of \citeres{'tHooft:1978xw,Beenakker:1988jr},
where different regularization schemes are translated into each other
as described in \citere{Dittmaier:2003bc}.  Our treatment of rational
terms of UV or IR origin is described in Appendix~A of
\citere{Bredenstein:2008zb}.  More details on the two independent
calculations of the virtual corrections in the gg channel will be
presented elsewhere.

In both evaluations of the real corrections the amplitudes are
calculated in the form of helicity matrix elements which have been
generated with {\sc Madgraph 4.1.33} \cite{Alwall:2007st}. While the
amplitudes for $q\bar q\to\Pt\bar\Pt\Pb\bar\Pb\Pg$ have been checked
with the spinor formalism of \citere{Dittmaier:1998nn}, those for
$\Pg\Pg\to\Pt\bar\Pt\Pb\bar\Pb\Pg$ have been verified with an
implementation of off-shell recursion relations
\cite{Berends:1987me,Caravaglios:1995cd,Draggiotis:1998gr}. The
singularities for soft or collinear gluon emission are isolated via
dipole
subtraction~\cite{Catani:2002hc,Catani:1996vz,Dittmaier:1999mb,Phaf:2001gc}
for NLO QCD calculations using the formulation~\cite{Catani:2002hc}
for massive quarks.  
One of the two calculations employs the automatic 
{\sc MadDipole} implementation of dipole subtraction~\cite{Frederix:2008hu}.
After combining virtual and real corrections,
singularities connected to collinear configurations in the final state
cancel for ``collinear-safe'' observables after applying a jet
algorithm. Singularities connected to collinear initial-state
splittings are removed via $\overline{\mathrm{MS}}$ QCD factorization
by PDF redefinitions.  In both evaluations the phase-space integration
is performed with multichannel Monte Carlo
generators~\cite{Hilgart:1992xu} and adaptive weight optimization
similar to the one implemented in {\sc
  Lusifer}~\cite{Dittmaier:2002ap}.


In the following we consider the process $\Pp\Pp\to\Pt\bar\Pt\Pb\bar\Pb+X$ at the LHC,
i.e.\ for $\sqrt{s}=14\TeV$.  For the top-quark mass, renormalized in
the on-shell scheme, we take the numerical value $\Mt=172.6\GeV$
\cite{Group:2008nq}. All other QCD partons (including $\Pb$~quarks)
are treated as massless particles, and collinear final-state
configurations, which give rise to singularities, are recombined into
IR-safe jets using a \mbox{$k_{\rT}$-algorithm} \cite{Catani:1992zp}.
Specifically, we adopt the \mbox{$k_\rT$-algorithm} of
\citere{Blazey:2000qt} and recombine all final-state b quarks and
gluons with pseudorapidity $|\eta| < 5$ into jets with separation
$\sqrt{\Delta\phi^2+\Delta y^2}>D=0.8$ in the
rapidity--azimuthal-angle plane.  Requiring two b-quark jets, this
also avoids collinear singularities resulting from the splitting of
gluons into (massless) b quarks.  Motivated by the search for a
$\Pt\bar \Pt H (H\to\Pb\bar \Pb)$ signal at the LHC
\cite{Cammin:2003,Cucciarelli:2006}, we impose the following
additional cuts on the transverse momenta, the rapidity, and the
invariant mass of the two (recombined) b jets:
\mbox{$p_{\rT,\Pb}>20\GeV$}, \mbox{$|y_\Pb|<2.5$}, and
$m_{\Pb\bar\Pb}> m_{\Pb\bar\Pb,\mathrm{cut}}$. We plot results either
as a function of $m_{\Pb\bar\Pb,\mathrm{cut}}$ or for
$m_{\Pb\bar\Pb,\mathrm{cut}}=0$.  Note, however, that the jet
algorithm and the requirement of having two b~jets with finite
$p_{\rT,\Pb}$ in the final state sets an effective lower limit on the
invariant mass $m_{\Pb\bar\Pb}$ of roughly $20\GeV$.  The outgoing
(anti)top quarks are neither affected by the jet algorithm nor by
phase-space cuts.

We consistently use the CTEQ6~\cite{Pumplin:2002vw,Stump:2003yu} 
set of PDFs, i.e.\ we take CTEQ6L1 PDFs with a
1-loop running $\alpha_{\mathrm{s}}$ in LO and CTEQ6M PDFs with a
2-loop running $\alpha_{\mathrm{s}}$ in NLO, but the suppressed
contributions from b~quarks in the initial state have been neglected.
The number of active flavours is $N_{\mathrm{F}}=5$, and the
respective QCD parameters are $\Lambda_5^{\mathrm{LO}}=165\MeV$ and
$\Lambda_5^{\overline{\mathrm{MS}}}=226\MeV$.  In the renormalization
of the strong coupling constant the top-quark loop in the gluon
self-energy is subtracted at zero momentum. In this scheme, the running
of $\alpha_{\mathrm{s}}$ is generated solely by the contributions of
the light-quark and gluon loops.
By default, we set the renormalization and factorization scales,
$\mu_{\mathrm{R}}$ and $\mu_{\mathrm{F}}$, to the common value
$\mu_0=\Mt+m_{\Pb\bar\Pb,\mathrm{cut}}/2$.

\begin{figure}
\includegraphics[bb= 95 445 280 655, scale=1.0]{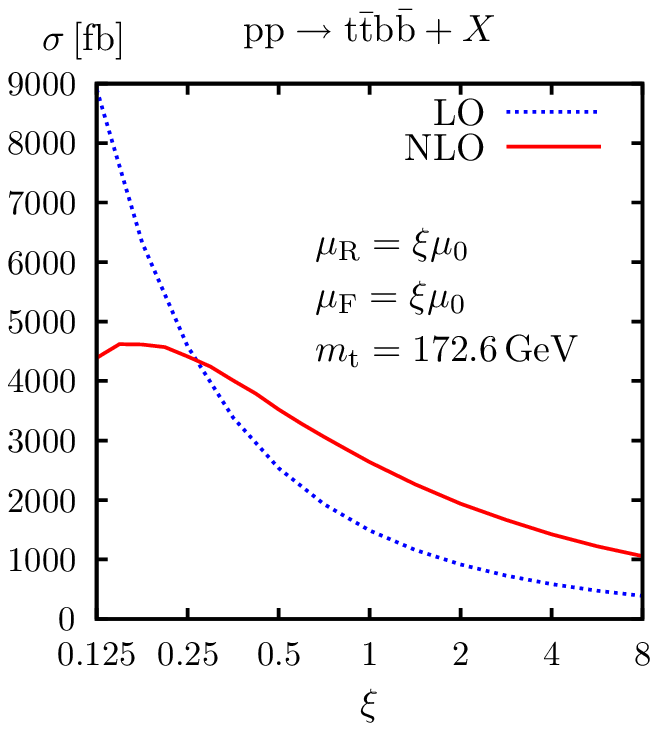}
\\[1ex]
\includegraphics[bb= 95 445 280 655, scale=1.0]{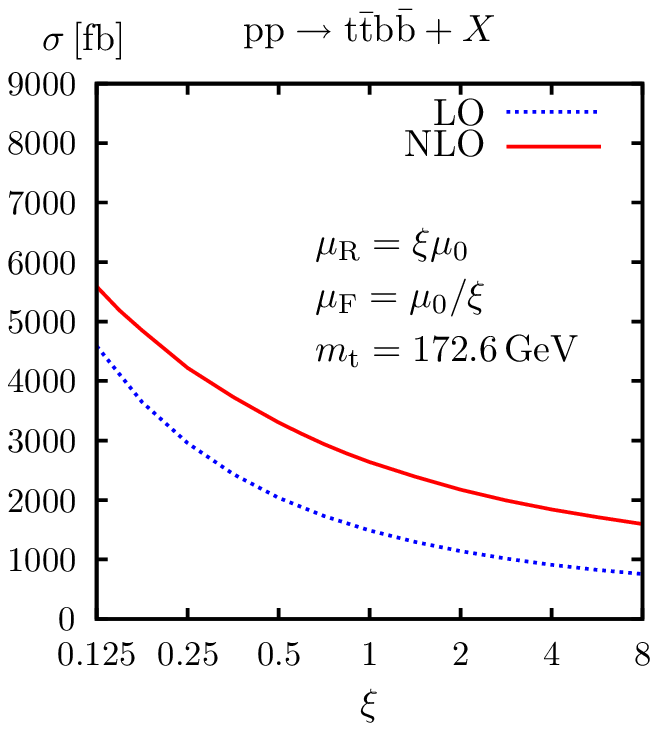}
\vspace*{-.8em}
\caption{Dependence of the LO and NLO cross sections of
$\Pp\Pp\to\Pt\bar\Pt\Pb\bar\Pb+X$ at the LHC
for $m_{\Pb\bar\Pb,\mathrm{cut}}=0$ and $\mu_0=\Mt$.}
\label{fig:scaledep}
\end{figure}

In \reffi{fig:scaledep} we show the scale dependence of the LO and NLO
cross sections upon varying the
renormalization and factorization scales in a uniform or an antipodal way.
We observe an appreciable reduction of the scale uncertainty upon
going from LO to NLO. Varying the scale up or down by a factor 2
changes the cross section by 70\% in LO and by 34\% in NLO.  At the
central scale, the full $\Pp\Pp$ cross section receives a very large
NLO correction of $77\%$, which is mainly due to the gluonic initial
states. Introducing a veto on extra jets by requiring
$p_{\rT,\mathrm{jet}}<50\GeV$ reduces the $K$-factor
to roughly 1.2.  For the $q\bar q$ channel we found a
very small correction of $2.5\%$~\cite{Bredenstein:2008zb}.  The full
LO and NLO cross sections are given by
$\sigma_{\mathrm{LO}}=1488.8(1.2)\fb$ and
$\sigma_{\mathrm{NLO}}=2638(6)\fb$, where the numbers in parentheses
are the errors of the Monte Carlo integration for $2\times10^7$
events.

\begin{figure}
\includegraphics[bb= 95 445 280 655, scale=1.0]{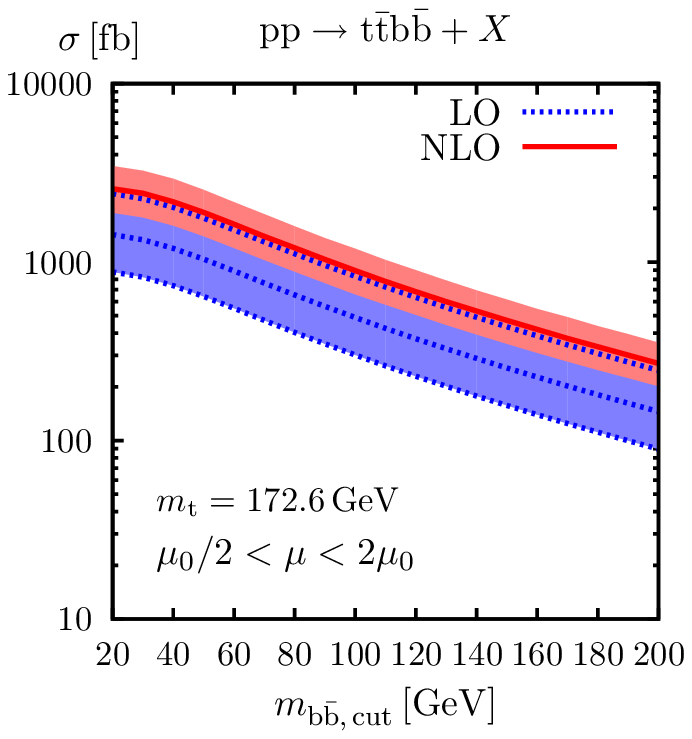}
\vspace*{-.8em}
\caption{LO and NLO cross sections
  for $\Pp\Pp\to\Pt\bar\Pt\Pb\bar\Pb+X$ at the LHC as function of
  $m_{\Pb\bar\Pb,\mathrm{cut}}$, with the bands indicating the scale
  dependence by varying $\mu_{\mathrm{R}}$ and $\mu_{\mathrm{F}}$ by
  factors $1/2$ and $2$ in a uniform or antipodal way
  ($\mu_0=\Mt+m_{\Pb\bar\Pb,\mathrm{cut}}/2$).}
\label{fig:NLOcsmbb}
\end{figure}

Figure~\ref{fig:NLOcsmbb} shows the LO and NLO cross sections as
function of the cut $m_{\Pb\bar\Pb,\mathrm{cut}}$ on the $\Pb\bar\Pb$
invariant mass, where the bands indicate the effect from a uniform or
antipodal rescaling of $\mu_{\mathrm{R}}$ and $\mu_{\mathrm{F}}$ by
factors $1/2$ and $2$.
The shown LO and NLO bands overlap in the whole considered range
in $m_{\Pb\bar\Pb,\mathrm{cut}}$, which is motivated
by the search for a low-mass Higgs boson. 
In contrast to the pure $q\bar q$ channel~\cite{Bredenstein:2008zb},
the NLO corrections to the full $\Pp\Pp$ process induce
only a moderate distortion of the functional dependence on 
$m_{\Pb\bar\Pb,\mathrm{cut}}$.
The reduction of the scale uncertainty from about $\pm70\%$ to
$\pm33\%$ and the large impact of the NLO correction hold true for the
considered range in $m_{\Pb\bar\Pb,\mathrm{cut}}$.


In summary we have presented first results on the full NLO prediction for the
process $\Pp\Pp\to\Pt\bar\Pt\Pb\bar\Pb+X$ at the LHC.  The NLO
corrections appreciably reduce the unphysical scale dependence of the
LO cross section, but at the same time enhance the cross section by a
$K$-factor of about $1.8$ for the usual scale choice. This large
correction factor can be strongly reduced by imposing a veto on hard
jets. It will be interesting to see how these NLO results influence
the signal significance of $\Pt\bar\Pt$--Higgs production with a Higgs
boson decaying into a $\Pb\bar\Pb$ pair, to which direct
$\Pt\bar\Pt\Pb\bar\Pb$ production represents a serious background.

On the technical side the presented calculation constitutes the first
complete NLO prediction for a hadronic process of the type $2\to4$
particles.  Speed and stability of the evaluation show good
performance of the Feynman-diagrammatic approach that is augmented by
dedicated reduction methods of tensor loop integrals for exceptional
phase-space regions. 

On a single 3\,GHz Intel Xeon processor, the evaluation of the virtual
corrections for $\Pg\Pg\to\Pt\bar\Pt\Pb\bar\Pb$ (including sums over
colour and polarization states) takes about $160\,\mathrm{ms}$ per
phase-space point.  This remarkably high speed suggests that the
employed reduction method might turn out to be a very efficient tool
for various other multi-particle processes at the LHC.
\\[.8em]
Acknowledgement: We thank Thomas Hahn for technical help in
structuring the very long source code, as well as M.~Mangano and
C.~Papadopoulos for helpful discussions on off-shell recursion
relations.  This work is supported in part by the European Community's
Marie-Curie Research Training Network under contract
MRTN-CT-2006-035505 ``Tools and Precision Calculations for Physics
Discoveries at Colliders'' and the Japan Society for the Promotion of
Science.

\bibliography{ppttbb_let}

\end{document}